\documentclass[twoside]{ilcws10}
\usepackage[latin1]{inputenc}
\usepackage[dvips]{graphicx,epsfig,color}
\usepackage{wrapfig,rotating}
\usepackage{amssymb,amsmath,array}

\newcommand{\tim}{$\times$}
\begin{document}
\title{
Status of Simulation Tools for the ILD  ScECAL} 
\author{Katsushige Kotera$^1$\footnote{Corresponding author (coterra@azusa.shinshu-u.ac.jp)}, 
Marc Anduze$^2$,
Vincent Boudry$^2$,
Jean-Claude Brient$^2$,\\
Daniel Jeans$^2$,
Kiyotomo Kawagoe$^3$,
Akiya Miyamoto$^4$,
Paulo Mora de Freitas$^2$,\\
Gabriel Musat$^2$,
Hiroaki Ono$^5$,
Tohru Takeshita$^1$,
 and 
Satoru Uozumi$^6$.
\vspace{.3cm}\\
1- Faculty of Science, Shinshu University, 
1-1-3 Asahi, Matsumoto, Nagano 390-8621, Japan\\
2- Laboratoire Leprince-Ringuet - \'{E}cole polytechnique, CNRS/IN2P3, Palaiseau, France \\
3- Department of Physics, Kobe University,
1-1 Rokkodai, Nada, Kobe, Hyogo 657-8501, Japan \\
4- High Energy Accelerator Research Organization (KEK), \\
1-1 Oho, Tukuba, Ibaraki 305-0801, Japan \\
5- Nippon Dental University School of Life Dentistry at Niigata,\\
 1-8 Hamaura-cho, Chuo-ku, Niigata 951-1500, Japan\\
6- WCU-High Energy Collider Physics Reseach, Kyungpook National University,\\
 1370 Sankyuk-dong, Buk-gu, Daegu 702-701, Korea 
}

\maketitle

\begin{abstract}
The scintillator-strip electromagnetic calorimeter (ScECAL) is one of the calorimeter technic for the ILC.
To achieve the fine granularity from the strip-segmented layers the strips in odd layers are orthogonal with respect to those in the even layers. In order to extract the best performance from such detector concept, a special reconstruction method and simulation tools are being developed in ILD collaboration.
This manuscript repots the status of developing of those tools.
\end{abstract}

\section{Introduction}

\begin{wrapfigure}[18]{r}{0.5\columnwidth}
\centerline{\includegraphics[width=0.42\columnwidth]{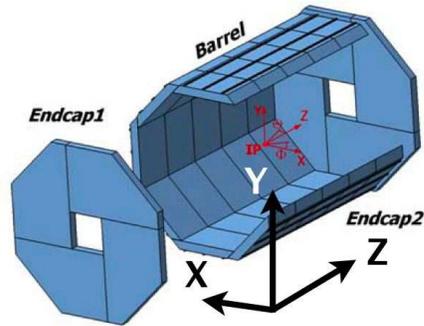}}
\caption{\small Global layout of the ECAL. Distance between two end-caps is 4900 mm long and the inner radius of the minimum inner radius is 1847.4 mm long. }\label{fig:geometry}
\end{wrapfigure}
In the ILD collaboration the particle flow calorimetry is the mandatory technique which requires at least
 10 mm \tim 10 mm lateral segmentation in the electromagnetic calorimeter (ECAL)  \cite{LOI}.
The scintillator-strip electromagnetic calorimeter (ScECAL) is one of the proposed concepts for such ECAL.
In the concept strip-shape scintillators in odd layers are put orthogonally with respect to those even layers 
and forms about 30 layers of sampling calorimeter.
The tungsten layers are put between
such sensitive layers as the absorber layers.
For example, 
a typical size of the scintillator strip in lateral dimension is 10 mm $\times$ 45 mm in order to achieve the effective
 10 mm \tim 10 mm segmentation, which has been implemented in the prototype module and tested at Fermi National Accelerator Laboratory (FNAL) in 2008 and 2009 \cite{FNALBT}.
To extract the best performance of such the strip segmentation detector, reconstruction algorithm is being developed. 
Including the development of such algorithm, status of the simulation tools for the study of 
the ScECAL performance is reported in this manuscript.
Resent studies suggest that 5 mm \tim\ 5 mm segmentation is better for ECAL \cite{LOI}. 
Therefore, 5 mm width scintillator strips are also discussed in this study.

\section{Framework}

\subsection{ScECAL in MOKKA}
\begin{wrapfigure}[32]{Hr}{0.5\columnwidth}
\centerline{\includegraphics[width=0.42\columnwidth]{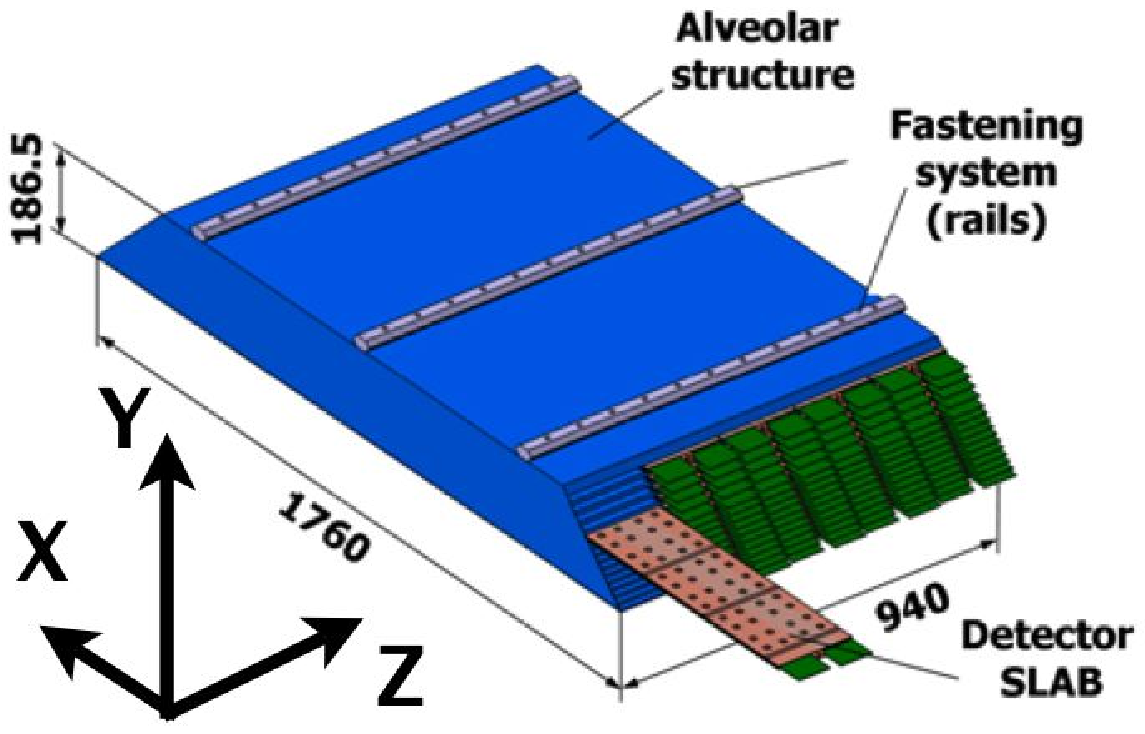}}
\caption{\small Layout of one module. A stave of the ECAL barrel is made with five modules.}\label{fig:module}
%
\centerline{\includegraphics[width=0.42\columnwidth]{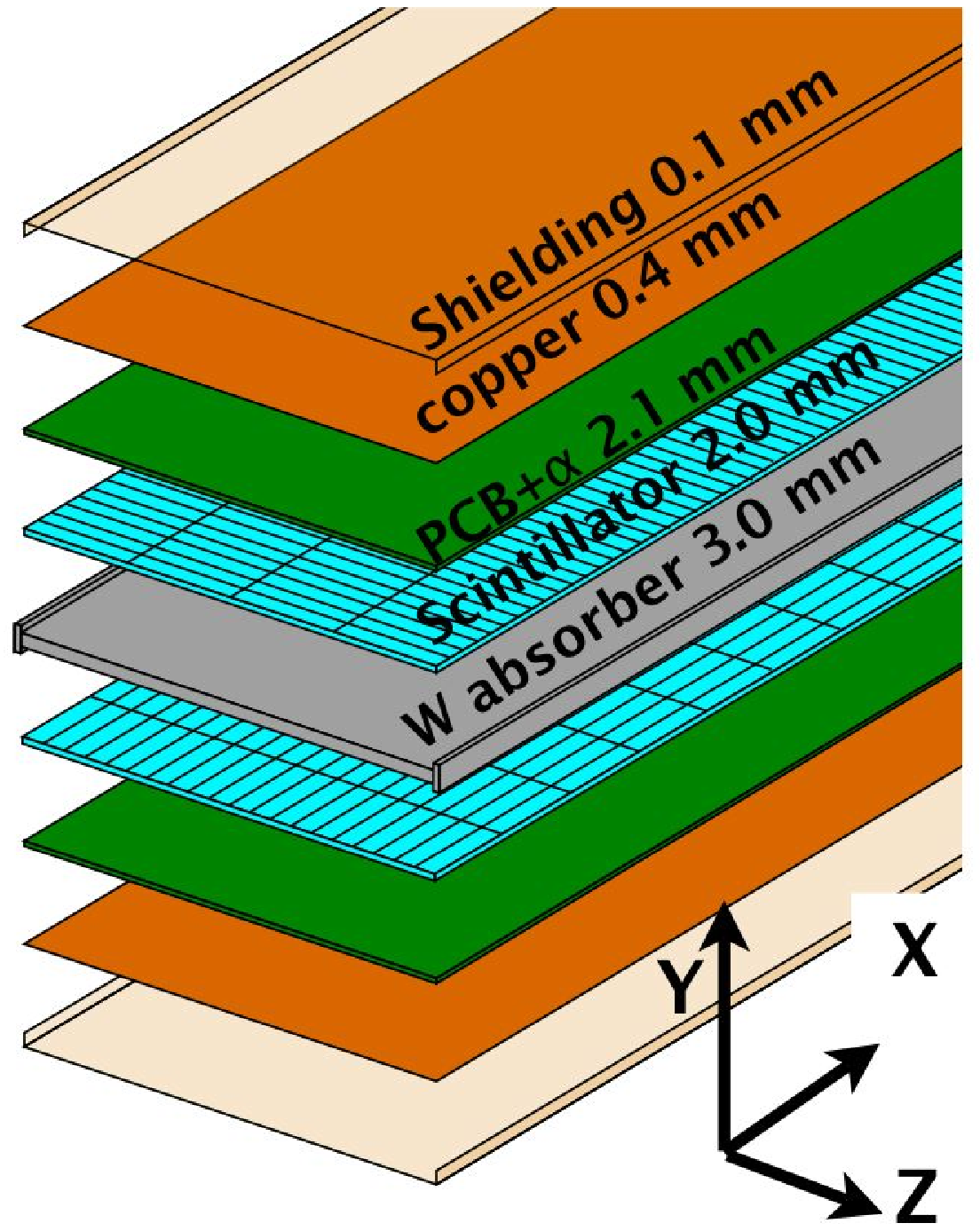}}
\caption{\small Structure of a unit of pair of ScECAL layers called ``SLAB''}\label{fig:slab}
\end{wrapfigure}

Figure \ref{fig:geometry} shows the geometry of the ECAL in MOKKA \cite{MOKKA} which is the detector simulator for the ILD.
The ScECAL has the same global structure as the Silicon tungsten ECAL (SiECAL) \cite{SIECAL} which is another concept of ECAL.
ECAL barrel part is made from eight staves 
in azimuthal angle direction and a stave is made of five modules as shown in Figure~\ref{fig:module}.

\noindent In the current plan the ScECAL is designed having 27 common thick layers,
while SiECAL has two different absorber thickness layers in 20 layers and 10 layers. 
A pair of layers where the scintillator strips in each layer are mutually orthogonal to the ones in the other layer,
constructs a detector unit called SLAB.
Figure~\ref{fig:slab} shows a structure of SLAB.
These two scintillator layers are mounted on the both above and below sides of a ``H'' shape tungsten absorber, and they 
are sealed between the printed circuit boards (PCBs).
Besides, the copper plates cover them as the thermal radiators.
With most outside shielding films, a SALB is inserted into a column of shelf structure made of the tungsten plates which also form absorber layers as shown in Figure~\ref{fig:module}.
The default thickness of the scintillator is 2.0 mm and tungsten absorber layers has 3.0 mm thickness. 
The hits in ECAL are generated as the hits on 5 mm \tim\ 5 mm square cells in MOKKA,
and hits on the square cells then are merged into strip shapes in the digitizer of 
the Marlin \cite{MARLINRECO}.

\subsection{Reconstruction using the Triplet method in the MarlinReco}
\begin{wrapfigure}[23]{Hr}{0.65\columnwidth}
\centerline{\includegraphics[width=0.42\columnwidth]{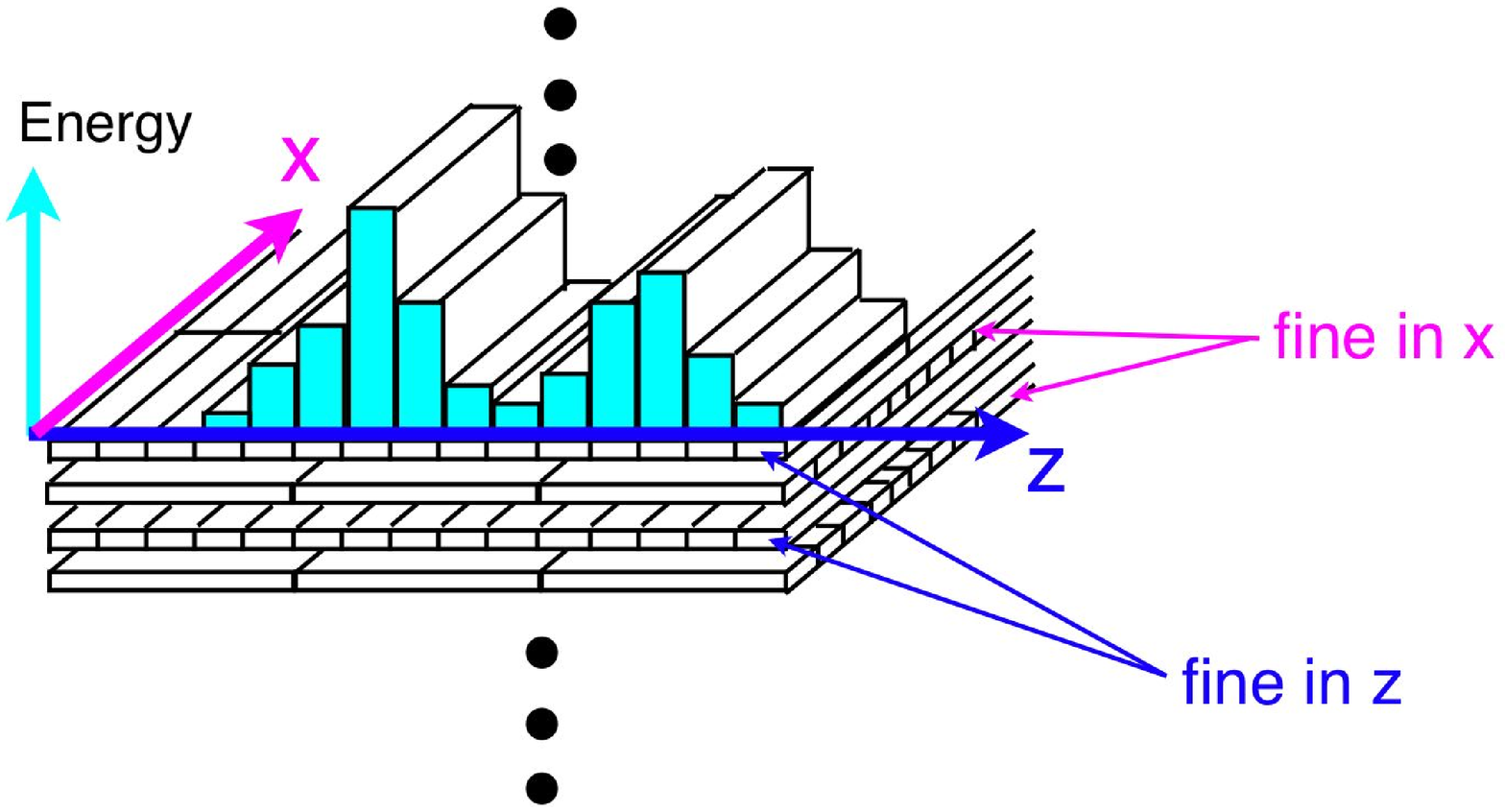}}
\caption{\small A schematic diagram of 2-D clustering }\label{fig:2dclustering}
\centerline{\includegraphics[width=0.6\columnwidth]{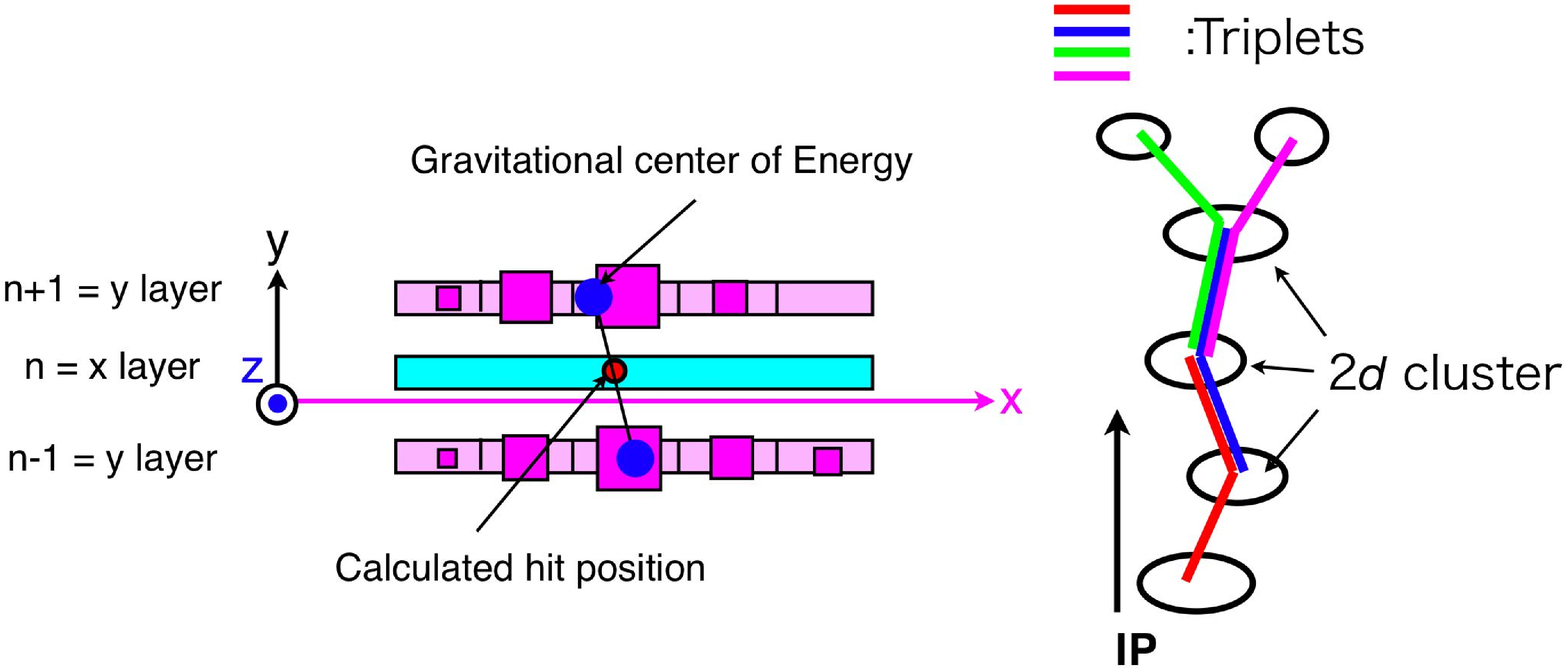}}
\caption{\small {\it Left}: Schematic diagram of triplet. {\it 
Right} combining the triplets which have two common 2-D clusters }\label{fig:triplet}
\end{wrapfigure}

To achieve the effective 5 mm $\times$ 5 mm segmentation with scintillator strips which have 5 mm width and some longer longitudinal length, the triplet method was developed by Daniel Jeans \cite{triplet}.
In the first step of the triplet method, the MIP tracks are searched considering only cells with a  hit consistent with the passage of a simple MIP through the cell, and which have a limited number of nearest neighbors in the same layer.
The TPC tracks associated with these calorimeter tracks are considered as 
the tracks break into the calorimeter.
Using the remaining unassociated hits 
the two dimensional calorimeter clustering (2-D clustering) is carried out.
The 2-D clustering is to make the clusters of neighbor strip in each layer.
When multiple energy peaks exist in a 2-D cluster, it is split into the clusters for the respective peaks.
Figure~\ref{fig:2dclustering} shows a schematic diagram of the 2-D clustering.
These 2-D clusters are combined into ``triplets'' in the next step.

\noindent A triplet is made from the three clusters in successive three layers, which overlap.
Figure~\ref{fig:triplet} left shows a schematic diagram to make a triplet, 
where the scintillator strips have longitudinal direction along the $x$ axis in $\mathbf{n}$ layer 
and they are finely segmented along to the $x$ axis in $\mathbf{n\pm1}$ layers. 
The hit position in the longitudinal direction of scintillator strip in $\mathbf{n}$ layer is calculated from the 
average of the energy weighted cluster position of above and below clusters.
After making triplets for every successive three layers, triplets which have two clusters in common are combined to make a "calorimeter track" explained in Figure~\ref{fig:triplet} right. This procedure is done from the inner surface of ECAL.
When a combination of calorimeter tracks reduces a discrepancy between the TPC track
momentum and calorimeter track energy,
such calorimeter tracks are combined together.

%
%
%

The position and deposited energy of the hits included in those calorimeter tracks are turned over to the PandoraPFAProcessor which makes the particle flow approach.
\begin{wrapfigure}[10]{r}{0.59\columnwidth}
\centerline{\includegraphics[width=0.6\columnwidth]{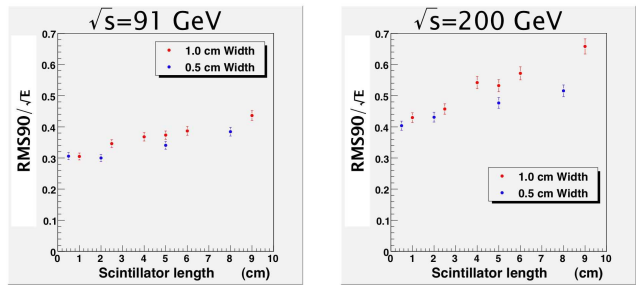}}
\caption{\small Jet energy resolution for the two Jets events of the 91 GeV center energy and 200 GeV center energy.}\label{fig:resolution}
\end{wrapfigure}
%
\section{Jet energy resolution}


Figure~\ref{fig:resolution} shows the jet energy resolution as a function of the length of 
the scintillator strips performed before the triplet method is implemented.
The energy resolution  gradually degrades as the length increases.
The triplet method is expected to reduce those degradation.

 %
\begin{wrapfigure}[40]{r}{0.4\columnwidth}
\centerline{\includegraphics[width=0.3\columnwidth]{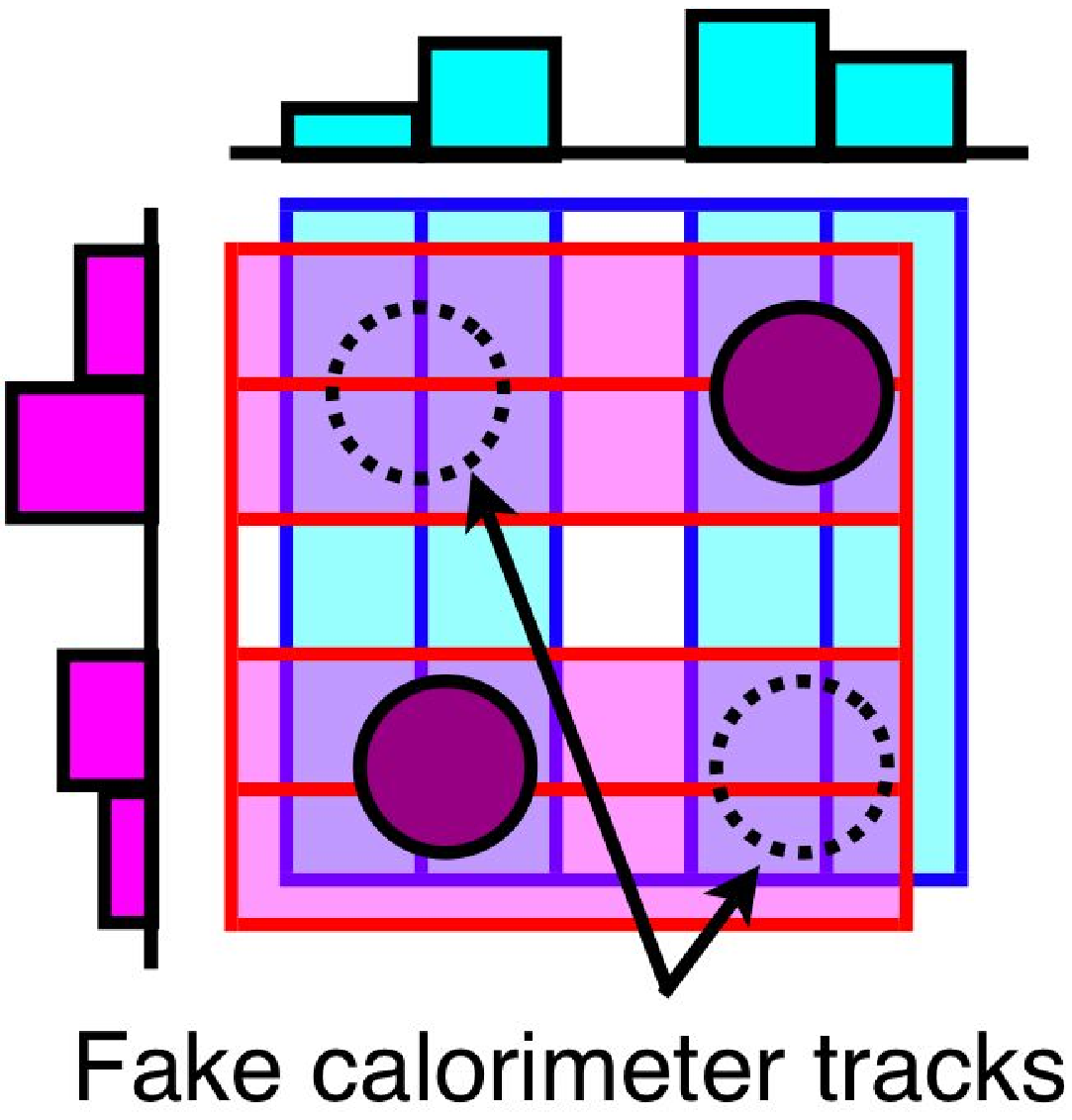}}
\caption{\small A case of two-fold ambiguity of calorimeter tracks.}\label{fig:twofoldambiguity}
\vspace{5mm}

\centerline{\includegraphics[width=0.3\columnwidth]{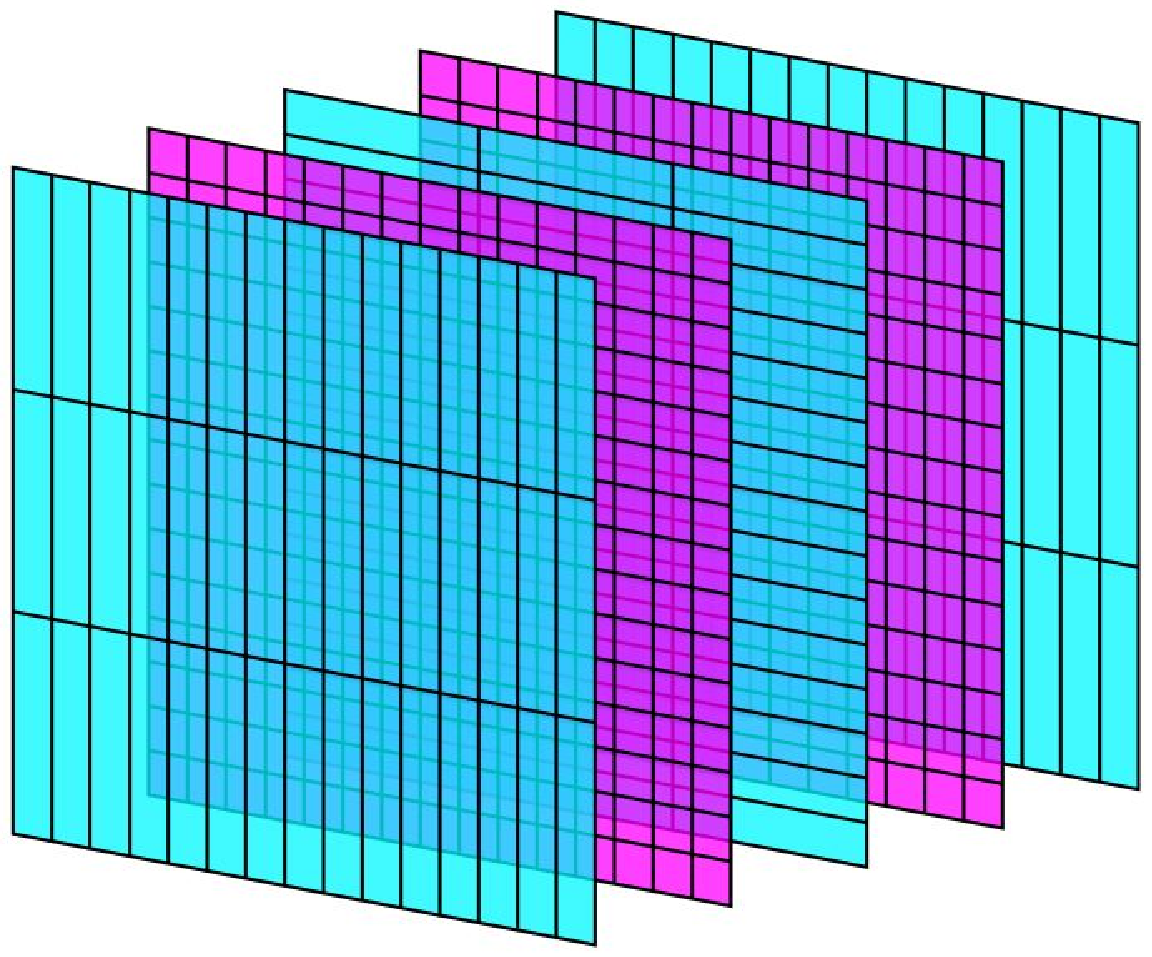}}
\caption{\small Hybrid of square cell layers and strip cell layers are suggested. This is one of the idea suggested. In the other case only first few layers are square cell layers.}\label{fig:hybrid}

\centerline{\includegraphics[width=0.40\columnwidth]{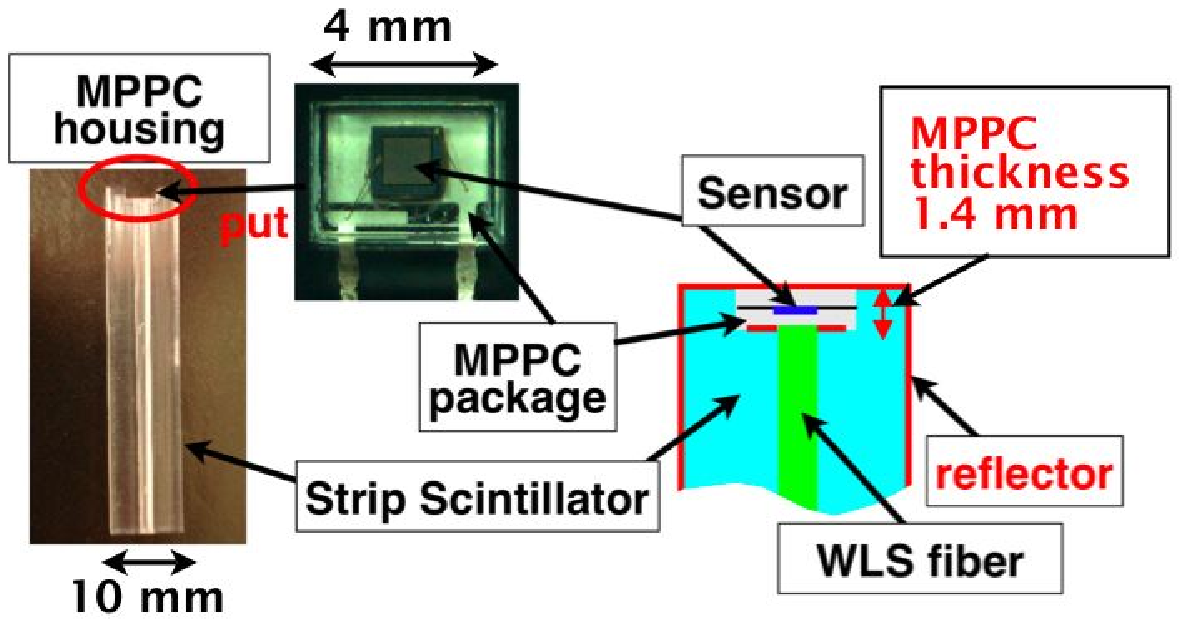}}
\caption{\small A typical scintillator strip unit tested at FNAL. MPPCs produced by Hamamatsu K.K. are used as PPDs}\label{fig:10x45strip}

\end{wrapfigure}
%

\section{Hybrid of the square cell layers and the strip cell layers}

One of the possible reason of the degradation of jet energy resolution
  is the two-fold ambiguity by the multi hits in a square which has four edges shaped with the longitudinal size of the scintillator strips shown in Figure\ref{fig:twofoldambiguity}.
 When the TPC track-information works correctly, the fake calorimeter tracks are merged into the correct calorimeter tracks optimizing the matching between TPC track momentum and the energy of the calorimeter track.
  
One of the way to resolve this problem is 
to replace some layers with some fine segmented square cell layers.
Figure~\ref{fig:hybrid} shows an example of such structure of ECAL layers.
We will study these type of detectors in future.


\section{Plan for the realistic simulation in MOKKA}

The smallest unit of sensitive scintillator detector is made of a plastic scintillator strip, a pixelated photon detector (PPD) sit in the housing made on the one side of the scintillator strip, wave length shift (WLS) fiber to collect the scintillation photons and a film reflector envelopes the scintillator. An example of the structure of the detector unit was used in the tested prototype module at  FNAL \cite{FNALBT}.
Figure~\ref{fig:10x45strip} shows the detector unit used in the prototype module.
Some of these materials make dead volume of detector. 
These dead volume potentially degrade the performance of the detector.
%
%

\begin{wrapfigure}[13]{r}{0.5\columnwidth}
\centerline{\includegraphics[width=0.4\columnwidth]{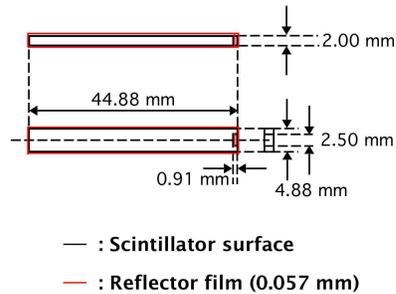}}
\caption{\small The new sensitive detector unit of the ScECAL.
This will be implemented in the next stage of MOKKA.}\label{fig:newDetail}
\end{wrapfigure}

According to our simulation study for the prototype module tested in FNAL, 
the effect from reflector film thickness is small enough on both the constant term and the stochastic term
of quadratic parameterization of the energy resolution.
Although the effect from the thickness of PPD is also small enough, the stochastic term of the energy resolution increase if the PPD thickness increases more than current size.
Additionally, we have a plan to change the width of the scintillator strip to 5 mm for effective segmentation 
of 5 mm $\times$ 5 mm. In this case the dead volume increases.
Therefore the realistic simulation including the dead volume mentioned above is required.
The scintillator size is shown in Figure~\ref{fig:newDetail}.
In order to decrease the dead volume, the smallest size of MPPC produced by Hamamatsu K.K. will be used; the size of its package is 2.0 mm \tim\ 2.5 mm \tim\ 0.9 mm.
According 
to the study of position dependence of the sensitivity of  5 mm \tim\ 45 mm \tim\ 3 mm scintillator,
we do not need WLS fiber to improve the uniformity of sensitivity for the 5 mm width scintillator strip.
Although  more detail study of the position dependence of sensitivity without WLS fiber is ongoing,
our plan is not to use the WLS fiber in the next step.  
ILD software group has plan to implement these realistic simulation in MOKKA near future.

%
 

\section{Summary}
The goal of ScECAL group of ILD is to achieve precise jet energy resolution using strip shape scintillators 
with the same performance as the one done with 5 mm \tim \ 5 mm square cell.
If it is achieved, the amount of photon sensors can be reduced. 
To achieve this, we are developing the algorithm to extract the best performance from the ScECAL.
To avoid the two-fold ambiguity, hybrid calorimeter of the square cell layers and the strip cell layers are 
suggested.
The improvement of the geometry model in MOKKA is in progress to include those new configurations.

\section{Acknowledgments}

The authors would like to thank to ILD group for many vital and important discussions.
The authors would also like to express thanks to the  CALICE collaboration for providing the indispensable
information of the experiment. 
This work is supported in part by the Creative Science Research Grants No.~18GS0202 of the Japan Society for Promotion of Science (JSPS).


\begin{footnotesize}


\end{footnotesize}


\end{document}